# WEAK FERROMAGNETISM IN MAGNETOELECTRICS LiCoPO$_4$ AND LiNiPO$_4$
*Weak ferromagnetism in LiCoPO$_4$ and LiNiPO$_4$*

Yu. Kharchenko and N. Kharchenko
*Institute for Low Temperature Physics & Engineering, National Academy of Sciences*
*Pr. Lenin 47, 61103 Kharkiv*
*Ukraine*
*ykharchenko@ilt.kharkov.ua*

M. Baran and R. Szymczak
*Institute of Physics, Polish Academy of Sciences*
*Al. Lotnikow 32/46, 02-668 Warsaw*
*Poland*

**Abstract**: The magnetic properties of the antiferromagnetic magnetoelectric single-crystals LiCoPO$_4$ and LiNiPO$_4$ were investigated using the SQUID-magnetometer. The magnetization was measured in a longitudinal experimental geometry (M ∥ **H**) along the crystallographic axes a, b and c. The carried out investigations have shown that these compounds exhibit weak ferromagnetism. The weak ferromagnetic (WFM) moment is directed along the antiferromagnetic axis in both crystals. But the temperature behaviours of the weak ferromagnetic moments in LiCoPO$_4$ and LiNiPO$_4$ are different. The weak ferromagnetic moment in LiCoPO$_4$ changes non-monotonically. The essential horizontal negative bias of the magnetic hysteresis loop is observed in LiCoPO$_4$.

## INTRODUCTION

The LiCoPO$_4$ and LiNiPO$_4$ crystals belong to the well-known family of orthorhombic antiferromagnetic (AFM) magnetoelectric crystals.[1,2] Though the physical properties of these crystals were investigated rather intensively (see, e.g., Refs. 3-15), it is impossible to consider, that their magnetic structures are established finally. The magnetoelectric,[5-7] magnetooptic,[8] and magnetic properties [9] have been revealed which do not agreed with the simple collinear antiferromagnetic arrangements of $Co^{2+}$ and $Ni^{2+}$ spins, found at the neutron diffraction studies.

In particular under a cyclic change of a magnetic field the magnetoelectric properties of the $LiNiPO_4$ and $LiCoPO_4$ crystals exhibit a hysteresis of the "butterfly" type,[5-7] which is characteristic for magnetoelectrics being magnetized in the presence of magnetic and electric fields. Hysteresis of this type can arise if only the crystal has a weak ferromagnetic moment or if the crystal magnetization in a magnetic field contains the contribution, quadratic in the magnetic field strength. In any case the hysteresis indicates on absence of anti-inversion among symmetry operations of magnetic point groups of these crystals. However, the spin arrangements established in neutron diffraction studies [3,4] were in accord with the genuine antiferromagnetic structures. With the purpose of search of the weak ferromagnetism we have carried out by means of a SQUID-technique detailed temperature and field researches of magnetization of the $LiCoPO_4$ and $LiNiPO_4$ crystals in depending on their magneto-thermal history.

## EXPERIMENTAL RESULTS AND DISCUSSION

Prof. H. Schmid of Geneva University provided the $LiNiPO_4$ and $LiCoPO_4$ single crystals for our studies. The experiments were carried out on the Quantum Design MPMS-5 magnetometer. The error of the temperature stabilization was not over 0.015 K. The magnetization was measured in a longitudinal experimental geometry (M || **H**) along the crystallographic axes *a, b* and *c*. Possible deviations of a field direction from crystallographic axes did not exceed 1 - 2 degrees. Prior to each measurement of the temperature dependence the sample was heated to a temperature of not less than 3 $T_N$ ($T_N$ is the Neel temperature) and then cooled to 5 K in the presence of a magnetic field $H_{FC}$. The direction of the field $H_{FC}$ was either parallel to the measuring field ($H_{+FC}$) or antiparallel to it ($H_{-FC}$). Temperature dependence of the spontaneous magnetic moment $M_{WFM}$ was determined as a difference of the experimental dependences ($M_{+FC}(T) - M_{-FC}(T))/2$. The experimental results have shown that both $LiCoPO_4$ and $LiNiPO_4$ exhibit a weak ferromagnetism. The temperature behaviours of the WFM moments in them are different.

# 1. LiCoPO$_4$

The WFM moment changes non-monotonically in LiCoPO$_4$ (Fig 1). A maximum is observed near T = ½ T$_N$ and its value is equal to 0.135 G. The values of spontaneous magnetization along other axes most likely are equal to zero, as the measured values (0.005 G and 0.001 G in maximum) are much less than magnetization along the *b* axis (0.135 G) and temperature behaviour of magnetization measured along directions near to the *a* and *c* axes completely repeats temperature behaviour of the magnetization measured along the *b* axis. Most likely the measured moments along these directions are connected with inexact orientation of the measurement directions with respect to the crystal directions. The maximal values of the measured magnetization along the *a* and *c* directions correspond to deviation of the measurement directions from the crystallographic axes *a* и *c* on 2 and 0.5 degrees.

There are also other interesting features of magnetic behaviour of this crystal. At a cyclic change of the magnetic field the essential horizontal negative bias of the magnetic hysteresis loop is observed (Fig. 2). The hysteresis loop is shifted in a direction opposite to a direction of the magnetic field in which the sample was cooled. As far as we know, the bias of magnetic hysteresis loop was observed only in the chemically non-uniform materials containing FM/AFM interfaces: in systems of FM clusters and small particles embedded in AFM material, in FM films deposited on AFM single crystals, in FM particles coated by AFM film, and in multilayer FM/AFM films.[16-18] The exchange interaction between magnetic ions of an anisotropic AFM material and a FM material in the AFM/FM interface causes the unidirectional magnetic anisotropy for the FM material that give rise to the exchange bias of a magnetic hysteresis loop.

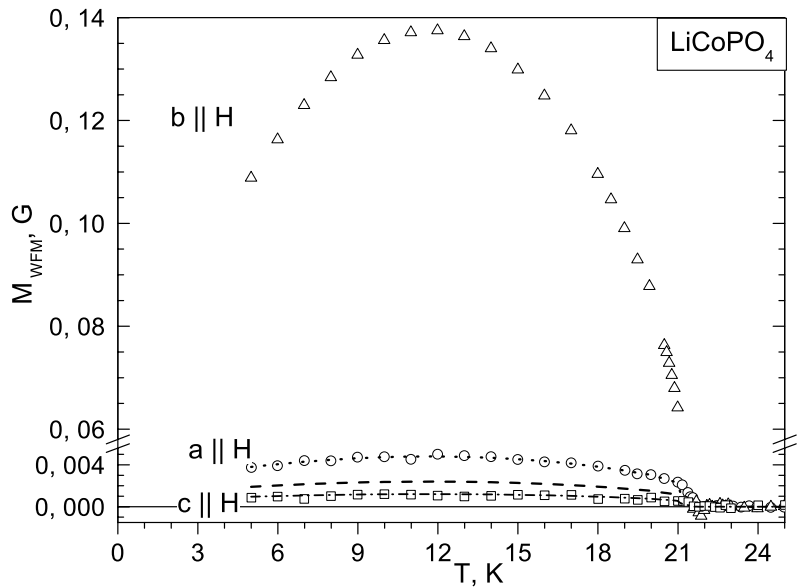

*Figure 1.* Temperature dependences of the spontaneous magnetization of LiCoPO$_4$ for directions close to *b*, *a,* and *c* (experimental points). Lines show the calculated values of magnetization along **a** or **c** axes under condition that a deviation of the WFM moment from the **b** axis is 2, 1, and 0.5 degrees (dot, dash, dash-dot) correspondingly.

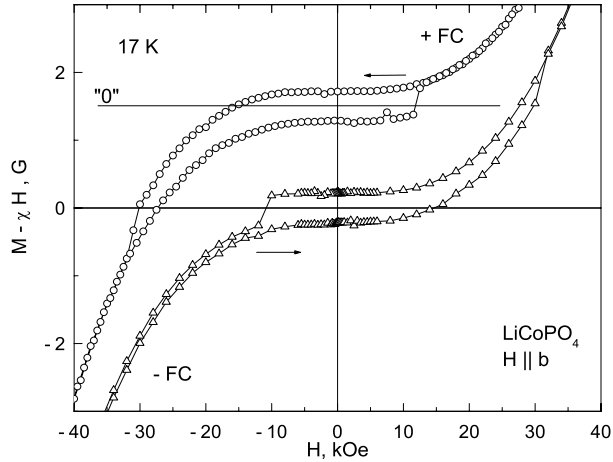

*Figure 2.* Magnetic hysteresis loops of the LiCoPO$_4$ single crystal: the M(H) curves after subtraction of the linear contribution $\chi H$ at T = 17 K after field cooling. The sample was cooled in field H$_{FC}$ = 1 kOe applied along the crystallographic *b* axes in two opposite directions.

In our case we deal with a stochiometric single crystal. The bias of the magnetic hysteresis loops observable in the pure single crystal LiCoPO$_4$ testifies that the magnetic structure of this antiferromagnetic crystal is not uniform. Obviously, only stratification of its magnetic structure into the WFM and AFM phase domains can cause the unidirectional anisotropy for WFM piece of the sample and cause a noticeable exchange bias of the magnetic hysteresis loop. The magnetic structure should be a modulated AFM/WFM structure, which at the low temperature is similar to the layered structure with the antiferromagnetic exchange links between the layers.

That fact, that sample has the residual (spontaneous) magnetic moment, indicates that the magnetic moment at modulation retains its direction and only its value varies. Besides, the basic mode L$_{2y}$ of the AFM ordering in LiCoPO$_4$ should remain uniform, as the crystal LiCoPO$_4$, cooled in a magnetic field, exhibits maximal magnetoelectric effect.[5,6] The completely modulated magnetic vectors can be only the transverse components of the basic AFM vector **L$_2$** and transverse components of others AFM vectors **L$_1$** and **L$_3$** of the minor AFM modes. Only the interactions which are described invariants not below than fourth order on spins can respond for WFM in LiCoPO$_4$.

Thermodynamic potential contains the symmetry-allowed gradient invariants L$_{1z}$(dL$_{2x}$/dx), L$_{2x}$(dL$_{1z}$/dx), L$_{1x}$(dL$_{2z}$/dx), L$_{2z}$(dL$_{1x}$/dx) and L$_{1x}$(dL$_{2x}$/dz), L$_{2x}$(dL$_{1x}$/dz), L$_{1z}$(dL$_{2z}$/dz), L$_{2z}$(dL$_{1z}$/dz) which allow a spatial modulation of the AFM components L$_{1z}$, L$_{2x}$ and L$_{1x}$, L$_{2z}$ along the *a* axes and L$_{1x}$, L$_{2x}$ and L$_{1z}$, L$_{2z}$ along the *c* axes. On the other hand there are the invariants M$_y$L$_{2y}$L$_{2x}$L$_{1z}$, M$_y$L$_{2y}$L$_{2z}$L$_{1x}$ and the gradient invariants M$_y$L$_{2y}$(dL$_{2x}$/dx)(dL$_{1z}$/dx), M$_y$L$_{2y}$(dL$_{2z}$/dx)(dL$_{1x}$/dx), M$_y$L$_{2y}$(dL$_{2x}$/dz)(dL$_{1z}$/dz), M$_y$L$_{2y}$(dL$_{2z}$/dz)(dL$_{1x}$/dz) which allow the WFM moment M$_y$. Certainly, there are also invariants P$_x$L$_{2x}$L$_{1z}$, P$_x$L$_{2z}$L$_{1x}$ permitting the spontaneous electric polarization of crystal along the x-axis. So, having supposed the noncollinear magnetic structure of crystal it is possible to see an opportunity of occurrence of the modulated weak ferromagnetism **M$_s$** || **b** in LiCoPO$_4$. One of possible mechanisms causing occurrence weak ferromagnetism can be the magnetoelectric mechanism.[7] In any case the crystal should have spontaneous very weak electric polarization in a direction of the c axis.

It is remarkable, that the non monotonous dependence of the WFM moment can be connected to evolution of the modulated structure from a nearly harmonious structure at high temperature to a soliton-like one at low temperature, as WFM magnetization should be proportional to a value of the gradient and volume of that part of the sample where the change of AFM vectors occurs.

It is necessary to notice that at reorientation of the WFM moment should occur the reorientation of all components of all AFM vectors. But in those parts of the crystal where the WFM moment is zero (or near zero) reorientation of the spins cannot occur and the AFM phase remains. For this reason the induced by magnetic field H<H$_{exch}$ (but not H>H$_{exch}$) remagnetized state of LiCoPO$_4$ is not equivalent to the homogeneous single domain state which is prepared by means of cooling of the sample in the magnetic field. This property of LiCoPO$_4$ should emerge experimentally at research of the effects sensitive to orientation of the AFM vector, for example, the magnetoelectric effect.

## 2. LiNiPO$_4$

The temperature dependence of the magnetization of the single crystal LiNiPO$_4$ is measured for magnetic-field orientation along the *a, b*, and *c* axes. It is found that the value of the magnetization depends on the magneto-thermal prehistory of the sample for **H** || **c**.

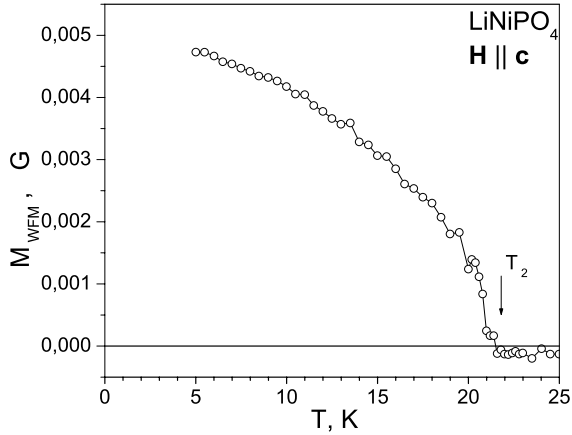

*Figure 3.* Temperature dependence of the spontaneous magnetization of the LiNiPO$_4$ single crystal. The magnetization was measured in field H=1 kOe, the sample was cooled in magnetic field H$_{+FC}$ = 15 kOe and in field H$_{-FC}$ = -15 kOe

The measured residual magnetization in LiNiPO$_4$ is much lower than in LiCoPO$_4$ and equal to near 0.005 G at 5 K (Fig. 3). The spontaneous moment is decreasing monotonically as the sample is heating, and it vanishes at a temperature near 20.8 K.

It is assumed that WFM in LiNiPO$_4$ is a property of the stochiometric crystal, then it as well as WFM in LiCoPO$_4$ can be described only by invariants not lower than fourth-order, since the usual invariants of the Dzyaloshinskii type are forbidden in the LiNiPO$_4$ and LiCoPO$_4$ antiferromagnets by the presence in the space symmetry group Pnma of an odd center of inversion with respect to the antiferromagnetic vector of main mode of ordering, **L$_2$ = S$_1$-S$_2$-S$_3$+S$_4$**.

Figure 4 shows the peculiarities of the temperature dependences of the magnetic susceptibility M(T)/H near the Neel temperature in magnetic fields **H** || a, **H** || b, and **H** || c. There are two features: a jump at T$_1$ and a kink at T$_2$. The observed jump cannot be due to the appearance of 180-degree AFM domains. Such domains can lead to vanishing of the magnetoelectric effect and the spontaneous magnetic moment, but they cannot be manifested in any way on the temperature dependence of the magnetic susceptibility. A jump of the susceptibility can be due solely to a jump in the AFM order parameter. The second feature, the kink, is observed as well for all three directions of the magnetic field. It is natural to suppose that at temperature T$_2$ a second-order phase transition occurs. In the absence of a magnetic field the temperature of these phase transitions are close to T$_1$=20.8(5) K and T$_2$=21.8(5) K. It should be mention that a smooth transition to the intermediate phase occurs on the high-temperature side, while the transition on the low-temperature side is abrupt. Such a sequence of transitions is characteristic for transitions to an incommensurate phase.[19] This circumstance suggests that the intermediate phase in LiNiPO$_4$ is the incommensurate AFM phase. The formation of a modulated phase in at temperature close to T$_N$ may facilitate a weak exchange interaction between the AFM layers and weak magnetic anisotropy in the *ac* plane at these temperatures.

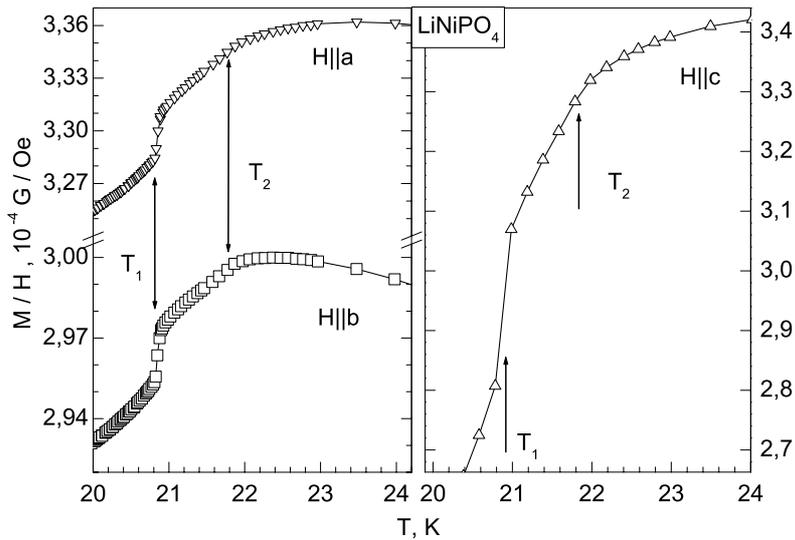

*Figure 4.* Temperature dependence of the magnetic susceptibility M(T)/H of LiNiPO$_4$ in an external magnetic field H = 20 kOe (**H || a, H || b**) and 10 kOe (**H || c**).

The results of these magnetic measurements of LiNiPO$_4$ agree in part with the results of neutron diffraction studies by D. Vaknin with co-authors presented on this Conference.

## CONCLUSION

The carried out investigations have shown that the magnetoelectric LiCoPO$_4$ and LiNiPO$_4$ crystals exhibit weak ferromagnetism. The essential horizontal negative bias of the magnetic hysteresis loop and the non-monotonically changes observed in LiCoPO$_4$ indicate that the WFM moment in this AFM crystal is the modulated one, and direction of the WFM moment is remained at the modulation. At low temperature the magnetic structure should be similar to a stratified structure of the AFM/WFM type. Only the interactions which are described by the invariants at least of fourth order on spins can be responsible for WFM in LiCoPO$_4$ and LiNiPO$_4$ crystals. One of the possible mechanisms for the appearance of weak ferromagnetism in these AFM crystals is magnetoelectric one.

The revealed weak ferromagnetism in the LiCoPO$_4$ and LiNiPO$_4$ crystals raises the question of redetermination of the magnetic structures of these crystals, as the magnetic symmetry groups found earlier correspond to the genuine antiferromagnetic structures and forbid weak ferromagnetism.